\documentclass[english,12pt]{article}

\usepackage{babel}
\usepackage{graphics}

\makeatletter

\usepackage{graphicx}
\makeatother

\begin{document}

\noindent{\large{\bf {Monopole in the Dilatonic Gauge Field Theory}}}

\vspace{1in}

\noindent{\em D.Karczewska, R.Manka} \\
\textit{\emph{{\small \noindent{Department of Astrophysics and Cosmology,}}\emph{ }\textit{\emph{Institute of Physics, University of Silesia,}}\emph{ }\textit{\emph{ Uniwersytecka 4, 40-007 Katowice, Poland.}}}}

\vspace{1in}

{\noindent{\small Abstract}}

{\small \noindent{A numerical study of coupled to the dilaton field, static, spherically symmetric monopole solutions inspired by the Kaluza-Klein theory with large extra dimensions are presented. The generalized Prasad-Sommerfield solution is obtained. We show that monopole may have also the dilaton cloud configurations.} 

\medskip

\section{Introduction}

Recently there has been considerable interest in the field theories with large
extra spacetime dimensions. In comparison to the standard Kaluza-Klein theory
these extra dimensions may be restricted only to the gravity sector of theory
while the Standard Model (SM) fields are assumed to be localized on the 4-dimensional
spacetime (Antoniadis, Arkani-Hamed, Dimopoulos and Dvali (1998); Antoniadis,
Arkani-Hamed, Dimopoulos and Dvali (1998); (Arkani-Hamed Dimopoulos and Dvali (1999)).
It is promising scenario from the phenomenological point of view because it
shifts the energy scale of unification from \( 10^{19}\,  \)GeV  o \( 10-100\,  \)TeV. 

The gauge field theory is extended by inclusion of the dilatonic field in such
of theories. Such fields appear also in a natural way in Kaluza-Klein theories (Appelquist,
Chodos and Freund, (1987)), superstring inspired theories (Witten
(1985); Ferrata, L\"{u}st and Teisen (1989)) and in theories based on the noncommutative
geometry approach (Chamseddine and Fr\"{o}hlich (1993) ). 

As previous studies have already shown the inclusion of a dilaton in a pure
Yang - Mills theory has consequences already at the classical level. In particular
the dilaton Yang - Mills theories possess 'particle - like' solutions with finite
energy which are absent in the pure Yang - Mills case. Analogous equations have
recently been obtained for the 't Hooft - Polyakov monopole model coupled to
the dilatonic field (Lavreashvili and Maison (1992)).

\section{The dilatonic gauge field theory}

Dilatons appear in the higher dimensional theory after the process of the spacetime
compactification. The main idea of the theory with large extra dimensions is
that the gravity is realized in the more dimensional spacetime (the bulk) while
the matter is confined in the four-dimemsional spacetime (the brane). To be
clear and simple, we will consider the six-dimensional gravity. Let us now consider
the action integral of Einstein-Yang-Mills-Higgs theory in the six-dimensional
spacetime: 
\begin{equation}
\label{a1}
{\mathcal{S}}=\int d^{6}x\sqrt{-g_{6}}L,
\end{equation}
 where \( g_{6}=det(g_{MN}) \) and \( M=\{\mu ,\, i\} \), \( N=\{\nu ,\, j\} \)
with \( x^{M}=\{x^{\mu },y^{i}\},\, i=1,2 \). The metrical tensor in the six-dimensional
spacetime can be written: 
\begin{equation}
\label{a2}
g_{MN}=\left( \begin{array}{cc}
e^{-2\xi (x)/f_{0}}\overline{g}_{\mu \nu } & 0\\
0 & -r_{2}^{2}\delta _{ij}e^{+2\xi (x)/f_{0}}
\end{array}\right) ,
\end{equation}
 According to above definition we can write: 
\begin{equation}
\label{a3}
\sqrt{-g_{6}}=\sqrt{-\overline{g}}r_{2}^{2}e^{-2\xi (x)/f_{0}}.
\end{equation}
 In the equation (\ref{a2}) 
\begin{equation}
\label{g4}
g_{\mu \nu }=e^{-2\xi (x)/f_{0}}\overline{g}_{\mu \nu }
\end{equation}
 represents the four-dimensional metric in the Jordan frame while \( \overline{g}_{\mu \nu } \)
in the Einstein frame. We consider here the Lagrangian of the Einstein-Yang-Mills-Higgs
field as follows: 
\begin{eqnarray}
 & L=L_{g}+L_{YMH}\delta (y), & \\
 & L_{g}=-\frac{1}{2\kappa _{6}}(R-2\Lambda ), & \\
 & L_{YMH}=-\frac{1}{4}F_{\mu \nu }^{a}F^{a\mu \nu }+\frac{1}{2}D_{\mu }\Phi ^{a}D^{\mu }\Phi ^{a}-U(\Phi ), & 
\end{eqnarray}
 where \( \kappa _{6} \) is the six-dimensional gravitational coupling. \( L \),
\( L_{g} \), \( L_{YMH} \) - describe the total Lagrange function, the gravity
in six-dimensional spacetime and the Yang-Mills-Higgs field parts on the brane
emdeded in six-dimensional space, respectively. In general non-vanishing cosmological
constant (\( \Lambda \neq 0 \)) is possible. This case leads to the interesting
monopole solution (Lugo and Shaposhnik (1999, 2000)). In our paper we shall
focus our attention on the \( \Lambda =0 \) case. All calculations should include
metric \( g_{MN} \) (\ref{a3}), so for example \( D^{\mu }=g^{\mu \nu }D_{\nu } \).
Let us compactify the six-dimensional spacetime to the four-dimensional Minkowski
one on the torus \( ({\mathcal{M}}_{6}\rightarrow {\mathcal{M}}_{4}\times {\mathcal{S}}^{1}\times {\mathcal{S}}^{1}) \).
In this paper we assume that the extra dimensions are compactified to two-dimensional
torus with a single radius \( r_{2} \). The six-dimensional action may be rewritten
as: 
\begin{equation}
\label{a7}
{\mathcal{S}}=\int d^{4}x\int d^{2}y\sqrt{-g_{6}}L=\int d^{4}x\sqrt{-\overline{g}}{\mathcal{L}},
\end{equation}
 where \( \int d^{2}y=(2\pi r_{2})^{2} \) and \( {\mathcal{L}} \) is the effective
Lagrange function in four-dimensional spacetime. The six-dimensional gravitational
coupling \( \kappa _{6}=8\pi G_{6} \) is convenient to define as 
\[
G_{6}^{-1}=\frac{1}{(2\pi )^{2}}M^{4},\]
 where \( M \) - is the energy scale of the compactification (\( \sim 10-100\, TeV \)).
Compactification of the six-dimensional gravity on the torus gives the Lagrangian
(\ref{a7}) for the four-dimensional gravity
\begin{equation}
L=-\frac{1}{2\kappa }R_{(4)}
\end{equation}
 in the Einstein frame, where 
\begin{equation}
\label{aa8}
\frac{1}{\kappa }=\frac{(2\pi r_{2})^{2}}{\kappa _{6}}
\end{equation}
is the four-dimensional coupling constant or \( \kappa =8\pi G_{N}=8\pi M^{-2}_{Pl} \)
. From the above equation (\ref{aa8}) we get 
\begin{equation}
\label{mpl}
M^{2}_{Pl}=4\pi M^{4}r_{2}^{2}.
\end{equation}
 Cosmological consideration (Hall and Smith \textit{}(1999)) gives the bound
\( M\sim 100\, TeV \) which corresponds \( r_{2}\sim 5.1\times 10^{-5} \)
mm from equation (\ref{mpl}). Compactification of gravity on the five-dimensional
spacetime is rather unpysical (\( r_{2}\sim 10\, km \)), however the nice five-dimensional
spacetime compactivitation was propsed recently (Randal and Sundram (1999a,b)).
The Planck mass \( M_{Pl} \) (\ref{mpl}) is not longer a fundamental constant,
it may change during the Universe evolution (Flanagen Tye and Wasserman (1999)).
For four-dimensional Minkowski spacetime (\( \overline{g}_{\mu \nu }=\eta _{\mu \nu } \))

\begin{equation}
\label{r4}
R_{(4)}=\frac{4}{f_{0}^{2}}e^{2\xi (x)/f_{0}}\{-\partial _{\mu }\xi \partial ^{\mu }\xi +f_{0}\partial _{\mu }\partial ^{\mu }\xi \}.
\end{equation}
The last term in the equation (\ref{r4}) can by transform into the first one
by differentiating by parts. The parameter \( f_{0} \) (or re-scaling of the
\( \xi (x) \) field) is determined by the Planck mass (at present time) as:

\begin{equation}
f_{0}=\frac{1}{\sqrt{2\pi }}M_{Pl}\sim 4.87\, 10^{18}\, GeV/c^{2}
\end{equation}
 to produce the \( 1/2 \) term in for the dilaton field in (\ref{a9}). The
\( f_{0} \) parameter determines the dilaton scale \( f_{0} \). At the present
time \( f_{0} \) is rather high, so the interaction with dilatons can be neglected.
However, in the early universe when the Planck mass \( M_{Pl} \) was smaller
(for details see (Flanagen Tye and Wasserman (1999))) also the value of the
\( f_{0} \) was smaller, too. 

As a result of compactification of the six-dimensional Lagrangian we get the
Lagrange function for the Yang-Mills-Higgs fields. Fluctuations around the four-dimensional
Minkowski 
\begin{equation}
\overline{g}_{\mu \nu }=\eta _{\mu \nu }+h_{\mu \nu }(x,y)
\end{equation}
will produce interaction with Kaluza-Klein dilatons 
\begin{equation}
\label{grav}
h_{\mu \nu }(x,y)=\sum _{{\bf n}}h_{\mu \nu }^{{\bf n}}(x)e^{i\frac{2\pi n^{i}y^{i}}{r_{2}}}
\end{equation}
with the typical mass scale \( M \) (for \( n^{i}\neq 0 \)). 

In this paper we shall apply this approach to the simplest \( SO(3) \) gauge
field theory. The \( SO(3) \) gauge field theory has nice monopole solutions
('t Hooft (1976)) which have produced trouble in the cosmology and have been
the reason to introduce the idea of inflation. The main idea of this paper is
to examine how the monopole solution looks like in the dilatonic gauge field
theory inspired by the Kaluza-Klein gravity with the TeV scale. 

The dilatonic gauge field theory may described by the Lagrangian (defined by
equation (\ref{a7}), in the first approximation when \( \overline{g}_{\mu \nu }=\eta _{\mu \nu } \)
in the Minkowski spacetime) we get 
\begin{eqnarray}
 & {\mathcal{L}}=\frac{1}{2}\partial _{\mu }\xi (x)\partial ^{\mu }\xi (x)-\frac{1}{4}e^{2\xi (x)/f_{0}}F_{\mu \nu }^{a}F^{a\mu \nu }+ & \label{a9} \\
 & \frac{1}{2}(D_{\mu }\Phi ^{a})D^{\mu }\Phi ^{a}-e^{-2\xi (x)/f_{0}}U(\Phi ), & \nonumber \label{ref3} 
\end{eqnarray}
 where for \( SO(3) \) theory we have: 
\begin{equation}
\label{e1b}
U(\Phi )=\frac{\lambda }{4}(\Phi ^{a}\Phi ^{a}-v^{2})^{2}
\end{equation}
 with the \( SO(3) \) field strength tensor \( F_{\mu \nu }^{a}=\partial _{\mu }W_{\nu }^{a}-\partial _{\nu }W_{\mu }^{a}+g\epsilon _{abc}W_{\mu }^{b}W^{c}_{\mu } \).
The \( SO(3) \) gauge symmetry rotates the Higgs field \( \Phi ^{a} \). The
covariant derivative is given by \( D_{\mu }\Phi ^{a}=\partial _{\mu }\Phi ^{a}-g\epsilon _{abc}W_{\mu }^{b}\Phi ^{c} \).
Now we have \( D^{\mu }=\eta ^{\mu \nu }D_{\nu } \). The Higgs potential has
degenerate true vacua forming the sphere \( S^{2} \) \( (\Phi ^{a}\Phi ^{a}=v^{2}) \).
The Euler-Lagrange equations for the Lagrangian (9) are scale - invariant:\\

\begin{equation}
\label{t1}
x^{\mu }\rightarrow x^{\prime }{}^{\mu }=e^{\frac{u}{f_{0}}}x^{\mu },
\end{equation}

\begin{equation}
\label{t2}
\xi \rightarrow \xi ^{\prime }=\xi +u,
\end{equation}

\begin{equation}
\label{t3}
\Phi \rightarrow \Phi ^{\prime }=\Phi ,
\end{equation}

\begin{equation}
\label{t4}
W_{\mu }^{a}\rightarrow W_{\mu }^{a}=e^{-\frac{u}{f_{0}}}W_{\mu }^{a}.
\end{equation}
 These transformations change the Lagrange function in the following way: 
\begin{equation}
\label{t6}
{\mathcal{L}}\rightarrow {\mathcal{L}}^{\prime }=e^{-\frac{2u}{f_{0}}}{\mathcal{L}.}
\end{equation}
 This symmetry can be formulated equivalently as a scaling symmetry on the coordinates,
and the dilaton is often denoted as a Goldstone boson for dilatations. The origin
of the symmetry of the equations of motion is easy understood from the Kaluza-Klein
origin of the action. The scale transformations are equivalent to a rescaling
of the internal dimensions.

\section{The dilatonic monopole.}

The monopole scalar field configuration: 
\begin{eqnarray}
\Phi ^{a} & = & vh(r)n^{a}=vH(r)\frac{n^{a}}{gr},\label{fia} 
\end{eqnarray}
 where 
\begin{eqnarray}
n^{a} & = & \frac{x^{a}}{r}
\end{eqnarray}
\\
describes the 'hedgehog' structure \( n^{a} \) and scalar spherically symmetric
field \( H(r) \). The \( SO(3) \) gauge field is described by the \( K(r) \)
field: 
\begin{equation}
\label{fib}
A_{i}^{a}=-\varepsilon _{aij}\frac{1}{gr}n^{j}(1-K(r)).
\end{equation}
 The dilaton field is described by the \( S(x) \) function: 
\begin{equation}
\xi (x)=f_{0}S(x)
\end{equation}
 (if we introduce the dimensionless variable \( x=gvr \)). The field Euler-Lagrange
equations generated by Lagrange function (\ref{a9}) for \( H(x) \) and \( K(x) \)
and \( S(x) \) are: 
\begin{equation}
\label{eqa1}
H^{\prime \prime }(x)-\frac{2}{x^{2}}H(x)K(x)^{2}+\frac{\varepsilon }{x^{2}}e^{-2S(r)}(x^{2}-H^{2}(x))H(x)=0,
\end{equation}

\begin{equation}
\label{eqa2}
K^{\prime \prime }(x)+2K^{\prime }(x)S^{\prime }(x)-\frac{1}{x^{2}}e^{-2S(x)}H(x)^{2}K(x)-\frac{K(x)}{x^{2}}(K(x)^{2}-1)=0,
\end{equation}

\begin{eqnarray}
S^{\prime \prime }(x) & + & \frac{2}{x}S^{\prime }(x)-\alpha ^{-1}\frac{1}{x^{4}}e^{2S(x)}\{(1-K(x)^{2})^{2}+2x^{2}K^{\prime }(x)^{2}\}\label{eqa3} \\
 & + & \frac{1}{2}\varepsilon \alpha ^{-1}e^{-2S(r)}(1-\frac{H^{2}(x)}{x^{2}})^{2}=0.\nonumber 
\end{eqnarray}
 In the dilatonic monopole we have two independent dimensionless constants:

\begin{equation}
\varepsilon =\frac{\lambda }{g^{2}},
\end{equation}

\begin{equation}
\alpha =(\frac{f_{0}}{v})^{2}.
\end{equation}
 The mass (or the lowest energy) of the monopole in the rest frame is: 
\begin{equation}
M_{mon}=\frac{4\pi v}{g}\int \rho (x)x^{2}dx,
\end{equation}
 with the energy density given by: 
\begin{eqnarray}
\rho (x) & = & \frac{1}{2}\alpha S^{\prime }(x)^{2}+\frac{1}{2x^{4}}e^{2S(x)}\{(1-K(x)^{2})^{2}+2x^{2}K^{\prime }(x)^{2}\}\nonumber \\
 & + & \frac{1}{2x^{4}}\{2H(x)^{2}K(x)^{2}+(xH^{\prime }(x)-H(x))^{2}\}\label{rrl} \\
 & + & \frac{1}{4}\varepsilon e^{-2S(x)}(1-\frac{H^{2}(x)}{x^{2}})^{2}.\nonumber \label{rho} 
\end{eqnarray}
 Inside the monopole (according to the equations (\ref{eqa1}), (\ref{eqa2})
and (\ref{eqa3})) the asymptotical behaviour when \( x\rightarrow 0 \) is given
as: 
\begin{equation}
\label{ce1}
K(x)=1-tx^{2}+O(3),
\end{equation}

\begin{equation}
\label{ce2}
H(x)=ux^{2}+O(3),
\end{equation}

\begin{equation}
\label{ce3}
S(x)=a+bx^{2}+O(3),
\end{equation}
 where \textit{u, t, a} are local parameters and \( b \) must be determined
as:
\begin{equation}
\label{bb}
b=\frac{24e^{2a}t^{2}-\epsilon e^{-2a}}{12\alpha }.
\end{equation}
 Far from the monopole core, if \( r\rightarrow \infty \, (x\rightarrow \infty ) \),
both functions H(x) and K(x) should describe the normal vacuum (\( \Phi ^{a}\Phi ^{a}=v^{2} \))
with \( H\rightarrow x \) and \( K\rightarrow 0 \) according to the \( H,\, K \)
function definition (\ref{fia},\ref{fib}) remembering that \( x=gvr \). In
this limit the energy density (\ref{rrl}) has a simple limit
\[
\rho (x)=\frac{1}{2}\alpha S^{\prime }(x)^{2}+\frac{1}{2x^{4}}e^{2S(x)}.\]
 The monopole mass in this limit may be rewritten as:
\begin{eqnarray*}
 & M_{mon}=\frac{4\pi v}{g}\int \rho (x)x^{2}dx= & \\
 & \frac{4\pi v}{g}\int (\sqrt{\alpha }xS'+\frac{1}{x}e^{S})^{2}dx+\frac{4\pi v}{g}\sqrt{\alpha }(e^{S(0)}-e^{S(\infty )}). & \nonumber 
\end{eqnarray*}
The first term vanishes if the dilaton field obeys the Bogomolny equation 
\[
\sqrt{\alpha }xS'+\frac{1}{x}e^{S}=0.\]
 This equation has a nice solution in the uniform normal vacuum (Bizon (1993))
\begin{equation}
S_{b}=-ln((\sqrt{\alpha }e^{-S(\infty )}-1/x)/\sqrt{\alpha }).
\end{equation}
When \( r\rightarrow \infty  \) dilaton field should disappear in the true
vacuum. This demand gives \( S(\infty )=0 \). So, When \( r\rightarrow \infty  \)
we have the asymptotic behaviour of the solutions

\begin{equation}
\label{oo1}
H(x)=x\, (1-we^{-\sqrt{2\varepsilon }x})+O(1/x),
\end{equation}

\begin{equation}
\label{oo2}
K(x)=ze^{-x}+O(1/x),
\end{equation}
 
\begin{equation}
\label{oo3}
S(x)=-ln(\sqrt{\alpha }e^{-S(\infty )}-\frac{1}{\sqrt{\alpha }x})+O(1/x).
\end{equation}
 Even when \( S(\infty )\neq 0 \) it may be removed by the dilaton transformation
(\ref{t2}). We may solve the differential equations (\ref{eqa1},~\ref{eqa2},~\ref{eqa3})
by the iteration method expanding them with respect to \( \varepsilon  \):
\begin{equation}
H(x)=\sum ^{\infty }_{n=0}\varepsilon ^{n}H_{n}(x),
\end{equation}
 
\begin{equation}
K(x)=\sum ^{\infty }_{n=0}\varepsilon ^{n}K_{n}(x),
\end{equation}

\begin{equation}
S(x)=\alpha ^{-1}\sum ^{\infty }_{n=0}\varepsilon ^{n}S_{n}(x).
\end{equation}
In the first step \( (n=0) \) we obtain the equations: 
\begin{equation}
\label{pp1}
H^{\prime \prime }_{0}(x)-\frac{2}{x^{2}}H_{0}(x)K_{0}(x)^{2}=0,
\end{equation}

\begin{eqnarray}
\label{pp2}
 & K^{\prime \prime }_{0}(x)+\frac{2}{\alpha }K_{0}^{'}(x)S_{0}^{'}(x)-\frac{1}{x^{2}}e^{-2S_{0}(x)/\alpha }H_{0}(x)^{2}K_{0}(x) & \label{pp2} \\
 & -\frac{K_{0}(x)}{x^{2}}(K_{0}(x)^{2}-1)=0, & \nonumber  
\end{eqnarray}

\begin{equation}
\label{pp3}
S\prime \prime _{0}(x)+\frac{2}{x}S^{\prime }_{0}(x)-\frac{e^{2S_{0}(x)/\alpha }}{x^{4}}\{(1-K_{0}(x)^{2})^{2}+2x^{2}K^{\prime }_{0}(x)^{2}\}=0,
\end{equation}
leading to the Prasad-Sommerfield solution (without the dilaton field \( S(x) \)).
When \( \alpha \rightarrow \infty  \) we have the Prasad-Sommerfield solution
(Prasad and Sommerfield (1975)) and we can find for \( H_{0}(x) \) and \( K_{0}(x) \)
easily as :
\begin{equation}
\label{p1}
H_{0}(x)=x/tanh(x)-1,
\end{equation}
 
\begin{equation}
\label{p2}
K_{0}(x)=x/sinh(x).
\end{equation}
 Finding a nice analytical solution for the dilaton field (Fig.3, dotted line)
\begin{eqnarray}
\label{p3}
 & S_{0}(x)=a+\frac{Q_{D}}{x}+\frac{1}{4sinh(x)^{2}x^{2}}(-1+2x^{2}+2xQ_{D}+cosh(2x) & \label{p3} \\
 & -2xQ_{D}\, cosh(2x)-2x\, sinh(2x)) & \nonumber  
\end{eqnarray}
was a crucial point of this paper. The leading term for the dilaton field at
infinity will be a Coulomb one
\[
S_{0}(x)=\frac{Q_{D}}{x}+O(1/x)\]
 \( Q_{D} \) is the dilatonic charge which originates from the global scale
transformation (\ref{t1}-\ref{t4}). The similarity is striking, but we should
remember that an electric charge comes from the \( \exp (ie\alpha )\in U(1) \)
gauge symmetry. The global scale transformation (\ref{t1}-\ref{t4}) is generated
by the exponetrial transformation \( \exp (u/f_{0}) \). However, the asymptotic
bevaviour at \( x\rightarrow 0 \) (\ref{ce3}) admits \( Q_{D}=0 \). 

We present the numerical solutions of the coupled set of differential equations
(\ref{eqa1}, \ref{eqa2}, \ref{eqa3}) in the next section.

\section{Numerical solutions}

To solve the monopole equations numerically, we need the starting point (Press,
Teukolsky and Vetterlino (1992)). To find the starting conditions we can use
the solutions found from the variational procedures or from the Prasad-Sommerfield
approximation (\ref{p1}, \ref{p2}, 50). The trial solutions depending on the
variational parameters must be postulated in such a way to fulfill the boundary
conditions close to the center (\ref{ce1},\ref{ce2},\ref{ce3}) of the monopole
and at far outside (\ref{oo1},\ref{oo2},\ref{oo3}). We postulate the trial
solutions: 
\begin{equation}
\label{ev1}
H_{v}(x)=x\frac{(ux+x^{2}(1-e^{-\sqrt{2\varepsilon }x}))}{(1+x^{2})}\sim ux^{2}+O(3),
\end{equation}

\begin{equation}
\label{ev2}
K_{v}(x)=\frac{(1-tx^{2}+zx^{4})}{(1+x^{4}e^{x})}\sim 1-tx^{2}+O(3)
\end{equation}
 and 
\begin{equation}
\label{ev3}
S_{v}(x)=\frac{(a+bx^{2}+Q_{D}x^{5})}{(1+x^{6})}\sim a+bx^{2}+O(3),
\end{equation}
 where \( O(3) \) are corrections of the third order. For these functions the
monopole mass was calculated and the trial functions with minimal energy was
found. For the monopole without dilatons we get~ the \( M_{mon} \) mass of
monopole (if \( S(x)=0 \)):
\[
M_{mon}=15.886\times 10^{15}\, GeV.\]
 The trial functions give the dilaton configuration close to the monopole case
without dilatons. The minimal variational configurations for such a dilatonic
monopole for \( Q_{D}=0 \) and \( v=10^{16}\, GeV \) are presented on the
Table 1. This shows that the mass of the dilatonic monopole dependences on the
parameter \( \alpha  \) (\( f_{0} \)) and reaches the local minimum (for \( \alpha \sim 1 \))
lower than for monopole without dilatons.\begin{table}
{\centering \begin{tabular}{|c|c|c|c|c|}
\hline 
\( \alpha  \)&
u&
t&
z&
\( M_{dil} \) (10\( ^{15} \)GeV)\\
\hline 
\hline 
2.37\( \times  \)10\( ^{5} \)&
0.2396&
0.6021&
2.46903&
16.4125\\
\hline 
2.37\( \times  \)10\( ^{3} \)&
0.2396&
0.6021&
2.46886&
16.4107\\
\hline 
2.37&
0.2591&
0.5769&
2.2637&
15.0284\\
\hline 
1&
0.2879&
0.5389&
1.92221&
14.4402\\
\hline 
0.9&
0.2937&
0.5327&
1.8099&
14.4913\\
\hline 
0.8&
0.3011&
0.52609&
1.7035&
14.6313\\
\hline 
\end{tabular}\par}

\caption{The dependence of the monopole mass and the parameters u, t, z on the \protect\( \alpha \protect \)
parameter.}
\end{table}  

\begin{figure}
{\centering \resizebox*{12cm}{!}{\includegraphics{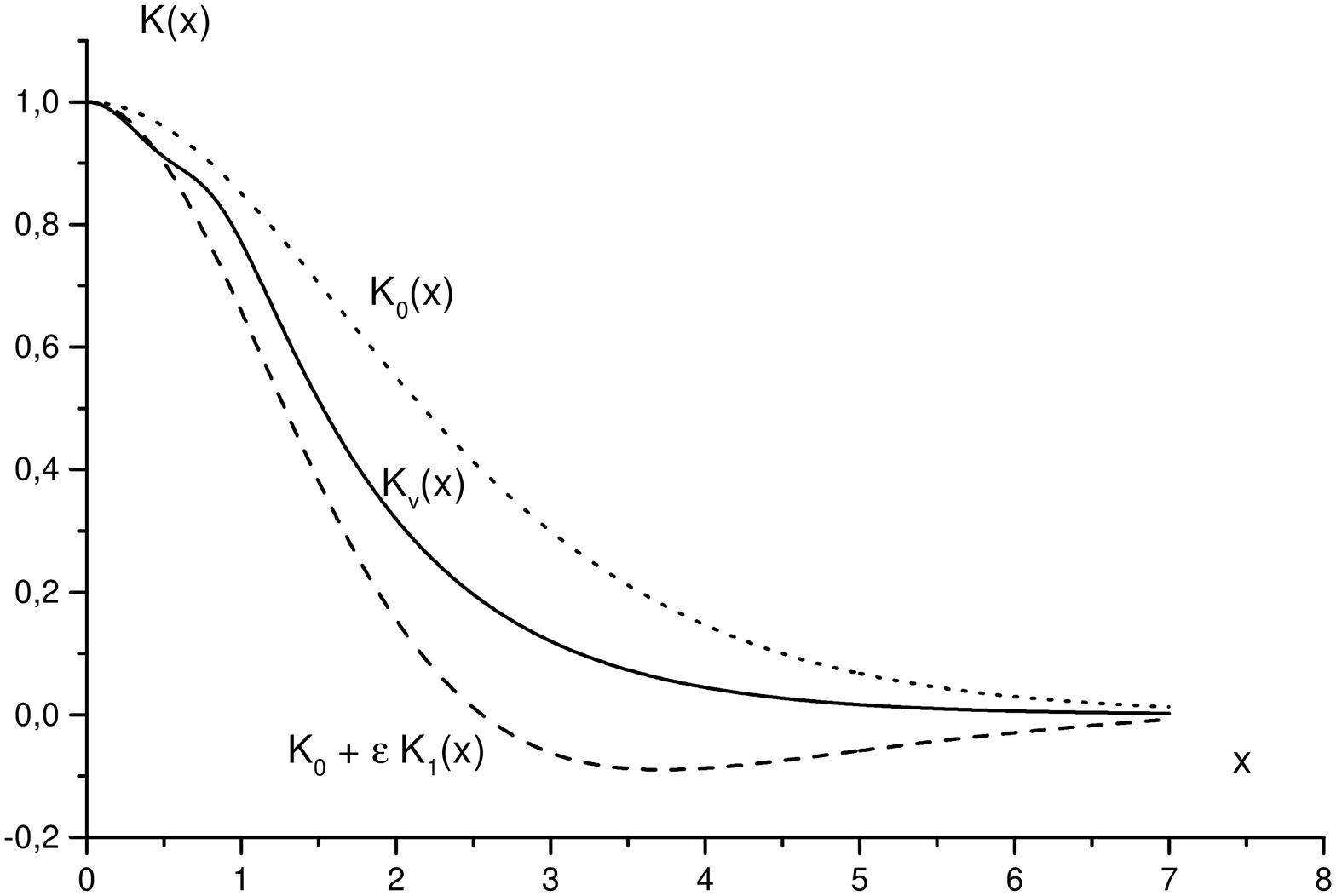}} \par}

\caption{The dependence of the gauge field \protect\( K(x)\protect \) on the \textit{x}
parameter. The dotted line corresponds to the Prasad-Sommerfield solution (\ref{p2}),
the solid one to the variational solution (\ref{ev2}), the dashed line presents
the solution of the first iteration.}
\end{figure}\begin{figure}
{\centering \resizebox*{12cm}{!}{\includegraphics{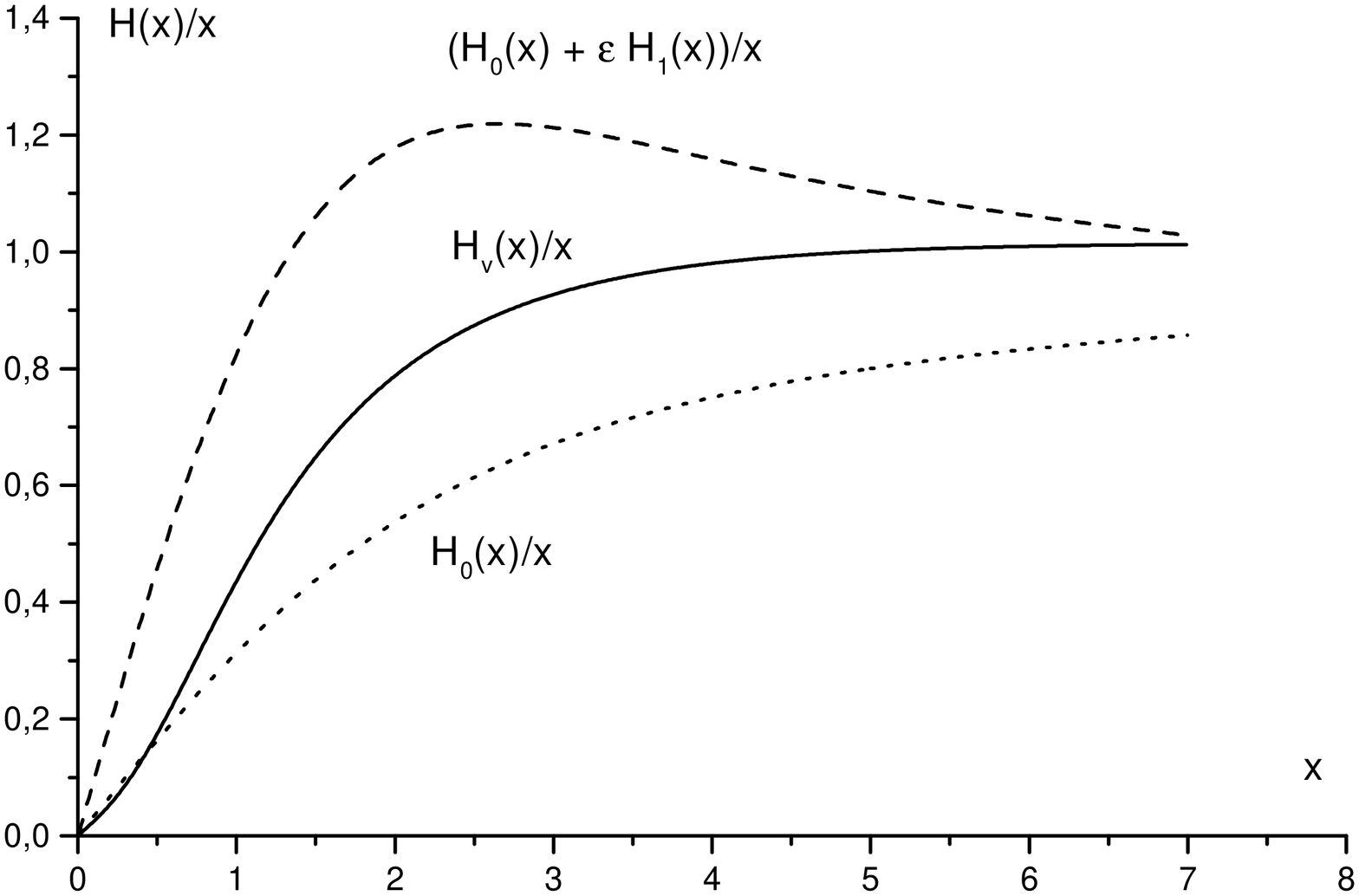}} \par}

\caption{The dependence of the Higgs field \protect\( H(x)/x\protect \) on the \textit{x}
parameter. The dotted line corresponds to the Prasad-Sommerfield solution (\ref{p1}),
the solid one to the variational solution (\ref{ev1}), the dashed line presents
the solution of the first iteration.}
\end{figure} 

For the dilatonic monopole the numerical method was independently verified using
the Chebyshev polynomial expansion (Michaila (1999)). The monopole solutions
for \( H(x) \) and \( K(x) \) are known very well so our attention is focused
on the dilaton solution \( S(x) \) especially. The behavior of the \( H(x) \)
and \( K(x) \) solutions determined by the boundary conditions is the same
as it is in the presence of the dilaton field. 

The Chebyshev method allows to calculate the exact solution of the differential
equations for the discrete set of points. The trial function provides the starting
data for the numerical solution of the ordinary differential equation (ODE)
(the shooting method (Press, Teukolsky and Vetterlino (1992) )) or the Chebyshev
functions method. After this preliminary numerical calculation the method based
on the Chebyshev polynomial was used. 

Clenshaw and Curtis have proposed almost forty years ago an integration method
based on the Chebyshev polynomials of the first kind of degree~\( j \), 
\begin{equation}
t_{j}(x)=cos(j*arccos(x)).
\end{equation}
 Since then, these methods have become standard. Since the Chebyshev polynomials
are orthogonal and allows to rewrite the function \( f(x) \) as 
\begin{equation}
f(x)=\sum _{j=0}\alpha _{j}t_{j}(x),
\end{equation}
 where (for \( j=0 \)) 
\begin{equation}
\alpha _{0}=\frac{1}{n}\sum ^{n}_{k=1}f(x_{k})t_{0}(x_{k}),
\end{equation}
 and (for \( j\neq 0 \)) 
\begin{equation}
\alpha _{j}=\frac{2}{n}\sum ^{n}_{k=1}f(x_{k})t_{j}(x_{k}).
\end{equation}
 The grid of \( n \) points \( x_{k} \) are zeros of the Chebyshev polynomial
\( t_{j}(x) \). This decomposition allows us to present the derivative of the
function \( f(x) \) as 
\begin{equation}
f^{\prime }(x_{i})=\sum _{k}D_{ik}f(x_{k}),
\end{equation}
 where the matrix 
\begin{equation}
D_{ik}=\sum ^{n-1}_{j=0}\frac{1}{c_{j}}t_{j}(x_{k})t_{j}^{\prime }(x_{i}),
\end{equation}
 and \( c_{0}=n \), \( c_{j}=n/2 \) (at \( j\neq 0 \)). This fact transforms
the ordinary differentional equation: 
\begin{equation}
\label{cz2}
\frac{d^{2}f}{dx^{2}}+p(x)\frac{df}{dx}+q(x)=r(x)
\end{equation}
 into an appropriate linear equation: 
\begin{equation}
\sum _{k}A_{ik}f(x_{k})=u_{i}.
\end{equation}
 So, we can have exact solution for discrete number of points. This method may
be also used to nonlinear equation 
\begin{equation}
\label{cz1}
m\frac{d^{2}f}{dx^{2}}+p(x)\frac{df}{dx}+F(f,x)=0.
\end{equation}
 If we have a starting function \( f_{0} \) then we expand \( f \) around
\( f_{0} \) 
\begin{equation}
f(x)=f_{o}(x)+\varepsilon (x),
\end{equation}
 and approximate the eq.(\ref{cz1}) with the eq.(\ref{cz2}) and then solve
numerically. The solution may be treat now as a starting function for the next
iteration, and so on. The iteration may last so long as an arbitrary precision
is reached.

The perturbation around \( \varepsilon  \) produces the series of differential
equations
\begin{equation}
f_{a,b}''+\sum _{b}p_{ab,n}(x)f_{b,n}'(x)+\sum _{b}q_{ab,n}(x)f_{b,n}(x)=r_{a,n}(x),
\end{equation}
where the vector 
\begin{equation}
f_{n}=\{H_{n}(x),\, K_{n}(x),\, S_{n}(x)\}.
\end{equation}
For example when \( n=1 \) the first equation (\( a=1 \)) corresponds to \( p_{1b}=0 \),
\( q_{11}=2K_{0}^{2}(x)/x^{2} \), \( q_{12}=2H_{0}(x)K_{0}(x)/x^{2} \), \( r_{1}=exp(-2S_{0}(x)/\alpha )(H_{0}^{2}(x)-x^{2})H_{0}(x)/x^{2} \),
and so on.

In the monopole case the starting function are these obtained by the variational
method. After expanding around trial functions (\ref{ev1},\ref{ev2},\ref{ev3})
we obtain system of the differential equations of the type (\ref{cz2}). After
that the numerical solution may be obtained on the grid of the \( x_{k} \). \begin{figure}
{\centering \resizebox*{12cm}{!}{\includegraphics{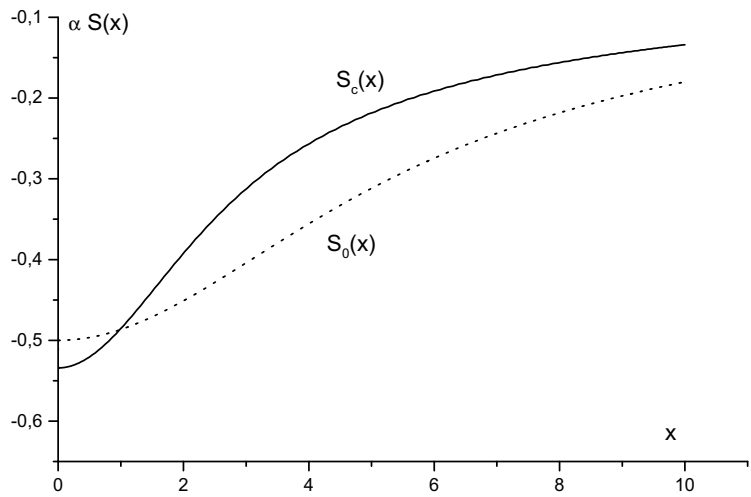}} \par}

\caption{The profile of the dilaton clound \protect\( S(x)\protect \). The solid line
presents the Chebyshev numerical solution \protect\( S_{c}(x)\protect \) and
the dotted line is the Prasad-Sommerfield solution \protect\( S_{0}(x)\protect \)
(50).}
\end{figure} The numerical solution for the dilaton field found by the Chebyshev numerical
method is presented on the Fig.3 (the solid line).

\section{Conclusions}

The aim of this paper was to present a numerical study of the classical monopole
solutions of the SO(3) theory coupled to the dilaton fields. We have shown that
a monopole is surrounded by the dilaton cloud \( S(x) \). In the field theories
with large extra spacetime dimensions the Planck mass is not longer a fundamental
constant and may change itself during the evolution of the universe. As a consequence
the parameter \( f_{0} \) changes, too. We have shown that the dilatonic monopole
reaches the minimal mass when \( f_{0}\sim v \) with the mass lower a bit than
for monopole without dilatons.

There is analytical solution in the Prasad-Sommerfield limit. 

The spherically symmetric dilaton solutions coupled to the gauge field or gravity
are interesting in their own and may moreover influence the monopole catalysis.

However in the theory inspired by the Kaluza-Klein theory with large extra dimensions
also new interaction with massive (\( \sim M \)) Kaluza-Klein gravitons takes
place. In the four dimensional spacetime the monopole solutions is stable due
to the monopole topological charge. Now the interaction with Kaluza-Klein gravitons
\( h_{\mu \nu }^{{\bf n}}(x) \) may cause disintegration of the monopole.

\subsection*{References }

{\small 

\noindent{Antoniadis, I., Arkani-Hamed, N., Dimopoulos, S. and Dvali, G., (1998) {\em Phys. Lett.} B {\bf 436}, 257}. 

\noindent{Appelquist, T., Chodos, A. Freund, P.G.O.(1987) {\emph{ Modern Kaluza-Klein Theories,} Meno Park (Addison-Wesley Publishing Comp.).} 

\noindent{Arkani-Hamed, N., Dimopoulos, S. and Dvali, G., (1998). {\em Phys. Lett.} B {\bf 429}, 263, }

\noindent{Arkani-Hamed, N., Dimopoulos, S. and Dvali, G., (1999) {\em Phys. Rev.} D {\bf 59}, 086004, (hep-ph/9807344) }

\noindent{Bizon, P. (1993). {\em Phys. Rev.} D, {\bf 47}, 1656. }

\noindent{Chamseddine, A.H., Fr\"{o}hlich, J. (1993). {\em Phys. Lett. } B {\bf 314}, 308 (hep-ph/9307209).}

\noindent{Ferrata, S., L\"{u}st, P. and Teisen, S., (1989). {\em Phys.Lett.} {\bf 233}, 147.}

\noindent{Flanagen, E.E., Tye S.-H., Wasserman I., {\em A Cosmology of the Brane World}, (hep-ph/9909373)}.

\noindent{Hall, L.J.,  Smith, D.,(1999). {\em Phys. Rev.} D {\bf 60} 085008 (hep-ph/9904267)};
Banks T., Nelson A., Dine M., (1999). {\em JHEP} {\bf 9906} 014, (hep-th/9903019). 

\noindent{Lavreashvili, G., Maison, D., (1992). {\em Phys. Lett.} {\bf 295}, 67; 
Lavreashvili, G., Maison, D., \textit{ Regular and black Hole Solutions of Einstein-Yang-Mills dilaton theory},preprint, MPI-ph/92-115(1992). preprint SLAC-PUB-7479, May 1997.}

\noindent{Lugo A. R., Shaposhnik F.A., (1999). {\em Phys Lett.} B {\bf 467} 43, (hep-th/9909226) }

\noindent{Lugo A. R., Moreno E.F., Shaposhnik F.A., (2000) {\em Phys. Lett.} B {\bf 473} 35 (hep-th/9911209)} \noindent{B.Michaila,\textit{ Numerical Approximation Using Chebyshev Polynomial Expansion}, (physics/9901005); 
Fox, I., (1962) {\em Computer Journal (UK)} {\bf 4}, 318, }; 
Clenshaw C.W., Norton H. J., (1963).{\em Computer Journal (UK)} {\bf 6}, 88. 

\noindent{Prasad, M.K., Sommerfield, C.M., (1975). {\em Phys.Rev.Lett.} {\bf 35}, 760. }

\noindent{Press, W.H., Teukolsky, S.A.,Vetterlino, W.T., {\it  Numerical Recipies: The art of Scientific Computing}, Cambridge University Press (1992). }

\noindent{Randal L., Sundram R., (1999a).{\em Phys. Rev. Lett} {\bf 83}, 3370, (hep-ph/9905221);}
Randal L., Sundram R., (1999b). {\em Phys. Rev. Lett} {\bf 83}, 4660, (hep-ph/9906064)

\noindent{'t Hooft, G., (1976). {\em Rev.Lett.} {\bf 37}, (1976), 11; {\em Phys.Rev.} D {\bf 14}, 3432.}

\noindent{Witten, E. (1985)\emph{.} {\em Phys.Lett.} B {\bf 245}, 561.}
}

\end{document}